\documentstyle[psfig,prb,aps]{revtex}
\begin{document}
\draft

\twocolumn[\hsize\textwidth\columnwidth\hsize\csname
@twocolumnfalse\endcsname

\title{
An {\em Ab Initio} Study of the Structures and Relative Stabilities of Doubly
Charged [(NaCl)$_m$(Na)$_2$]$^{2+}$ Cluster Ions
}
\author{Andr\'es Aguado$^*$}
\address{Physical and Theoretical Chemistry Laboratory, Oxford University,
South Parks Road, Oxford OX1 3QZ, UK.}
\maketitle
\begin{abstract}
We present {\em ab initio} perturbed ion calculations on the structures and
relative stabilities of doubly charged ((NaCl)$_m$(Na)$_2$)$^{2+}$
ions. The obtained stabilities show excellent agreement with experimental
abundances obtained from mass spectra. Those enhanced stabilities are found to
be a consequence of highly compact structures that can be built only for certain
values of $m$. Nearly all magic number clusters can be shown to be constructed
in one of the two following ways: (a) by adding tri- or penta-atomic chains
to two edges of a perfect neutral (NaCl)$_n$ cuboid, with $n$=$m$-2 or
$n$=$m$-4, respectively; (b) by removing a chloride anion from a perfect singly
charged (NaCl)$_n$Na$^+$ cuboid, with $n$=$m$+1.
\end{abstract}
\pacs{PACS numbers: 36.40.Wa; 36.40.Mr; 36.40.Qv; 61.46.+w}

\vskip2pc]

\section{Introduction}

Small alkali halide clusters have attracted in the last years the interest of 
both experimentalists and theoreticians because their simple ioniclike bonding
characteristics make them very easy to produce in pure form and also easily
amenable to theoretical modelling. They are ideal candidates, for example, to
study the behavior of solvated excess electrons and the insulator to metal
transition upon alkali enrichment in finite systems.
\cite{Yan92,Hon93,Hak94,Xia94,Lab95,Och95,Bon96,Fat97,Fra97,Dur97,Dur99,Ray00}
Structural isomerizations induced by a temperature increase,\cite{Che96,Doy99a}
as well as the finite system analogues of the bulk melting,
\cite{Ros92,Cal98,Doy99b} freezing,\cite{Ros93,Hua98} and glass\cite{Ros93b} 
transitions have also been studied. Apart from their inherent interest, it has
been shown that natural alkali halide clusters at the marine atmosphere can be
partially responsible for catalytic ozone depletion.\cite{Oum98}
All of these interesting properties of
alkali halide clusters largely depend on the specific structures adopted by 
them. Thus, a precise knowledge of the cluster structures is of paramount
importance. Very recently,
experimental techniques like electron diffraction from trapped clusters
\cite{Mai99} or measurements 
of cluster mobilities \cite{Mai97,Dug97,Hud97}
have been succesfully applied to study the structures of ionic 
clusters, and photoelectron spectroscopy has also been applied to study 
isomerization transitions in small alkali halide clusters.\cite{Fat96}
At the moment, however, these techniques need parallel theoretical
calculations to make a definite assignment of the observed diffraction pattern,
mobility or ionization potential to a specific isomer geometry.

The work on alkali halide clusters has been centered mostly in the singly
charged clusters (AX)$_n$A$^+$ and in the neutral clusters (AX)$_n$, where A is 
an alkali cation and X a halide anion. The abundance patterns obtained from the 
mass spectra of singly charged alkali halide cluster ions\cite{Cam81,Twu90} 
point towards a prompt establishment of bulk rock-salt symmetry. Theoretical
calculations have been able to rationalize the structures adopted by neutral
stoichiometric clusters in terms of cation/anion size ratios.
\cite{Agu97a,Agu97b} These studies show that small sodium iodide and lithium 
halide neutral
clusters adopt ground state structures based on the stacking of hexagonal rings,
while the rest of the materials adopt rocksalt-like ground state structures.
Theoretical calculations on the singly charged cluster ions are also available,
\cite{Agu98} and they conform to the experimental expectations of bulk rocksalt
symmetry even for those elements that crystallize in the CsCl-type structure,
namely CsCl, CsBr, and CsI. The emergence with increasing size of bulk CsCl 
structure in
(CsCl)$_n$Cs$^+$ cluster ions has been recently considered.\cite{Agu00}

The structural problem in neutral and singly charged alkali halide clusters
can thus be considered well understood. Work on doubly charged
[(AX)$_m$(A$_2$)]$^{2+}$ cluster ions has been much more scarce, probably due
to the inherent instability produced by the two excess charges. Sattler {\em et
al.}\cite{Sat81} published mass spectra of sodium iodide clusters, and
deduced a critical size for stability of the doubly charged series of $m$=18.
In the next year, Martin\cite{Mar82} reported pair potential model
calculations on the structures adopted by doubly charged sodium chloride
clusters, in an attempt to explain the experimental findings. He found that
all the cluster sizes had at least one metastable bound state, even though for
$m<$8 the removal of an Na$^+$ cation was an exothermic reaction. Li and
Whetten\cite{Li92} showed that there are important kinetic effects influencing
the critical size deduced by Sattler {\em et al,} and found that it is 
possible to populate the metastable minima for sizes smaller than that
critical size by using a less aggresive ionization technique. In that way they 
were able to find a ``stability island'' in the size region $m$=11--12. More
recently, Kolmakov {\em et al.}\cite{Kol99} have been able to observe such
small clusters as Cs$_5$I$_3^{2+}$ by embedding the alkali halide clusters
inside a rare gas coating that serves to dissipate the vibrational energy
excess adquired after ionization. Finally, Zhang and Cooks\cite{Zha00}
have published very recently mass spectra of free [(NaCl)$_m$(Na)$_2$]$^{2+}$
cluster ions in the size range $m$=11--62 by using electrospray ionization, and
found magic numbers for the sizes $m$=11,12,17,20,21,26,30,34,36,44,54, and 61.
They also report collision induced fragmentation spectra for those magic sizes,
from which a specific structural assignment is suggested. We would also like
to mention here the closely related case of metal oxide clusters, where only
one excess cation is needed to produce doubly charged isomers. Magnesium
and calcium oxide doubly charged cluster ions have been studied both
experimentally\cite{Mar89,Zie91,Zie92} and theoretically.\cite{Agu99,Agu00b}

To reach the same understanding level for the doubly charged clusters as that
already achieved for the singly charged ones, we present in this work
the results of an extensive and systematic theoretical
study of [(NaCl)$_m$(Na)$_2$]$^{2+}$ cluster ions with $m$ ranging from 6 to 28.
The experimental results by Zhang and Cooks\cite{Zha00} concerning enhanced 
stabilities will serve as an ideal check of the theoretical calculations. We
will show that we can reproduce all of their magic numbers, although we also
obtain some magic numbers not found in the experimental mass spectra. The
structural assignment suggested by these authors is also examined and shown to
be partially correct. The rest of the
paper is organized as follows: in Section II we give just a brief resume of the
theoretical model employed, as full an exposition has been reported already in
previous works\cite{Agu97a} and does not deserve in our opinion the use of more
journal space. The results are presented in Section III, and
the main conclusions to be extracted from our study in Section IV.

\section{The aiPI model. Brief location resume}

The {\em ab initio} Perturbed Ion (aiPI) model provides a computational
framework ideally suited to deal with ionic systems, and its performance has
been well tested both in the crystal\cite{Mar97,Agu98b,Agu99b,Agu00c} and 
cluster\cite{Agu97a,Agu97b,Agu98,Agu00,Agu99,Agu00b,Agu97c} limits.
The theoretical foundation of the aiPI model
\cite{Lua90} lies in the theory of electronic separability.\cite{McW94,Fra92}
Very briefly, the HF equations of
the cluster are solved stepwise, by breaking the cluster wave
function into local group functions (ionic in nature in our
case). In each iteration, the total energy is minimized with respect to
variations of the electron density localized in a given ion, with the electron
densities of the other ions kept frozen. In the subsequent iterations each 
frozen ion assumes the role of nonfrozen ion. When the self-consistent process 
finishes,\cite{Agu97a} the outputs are the total cluster energy and a set of 
localized wave functions, one for each geometrically nonequivalent ion of the 
cluster. These localized cluster-consistent ionic wave functions are then used 
to estimate the intraatomic correlation energy correction through Clementi's
Coulomb-Hartree-Fock method.\cite{Cha89,Cle00} The large multi-zeta basis sets 
of Clementi and Roetti\cite{Cle74} are used for the description of the ions. 
At this respect, our optimizations have been performed using basis sets (5s4p) 
for Na$^+$ and (7s6p) for Cl$^-$, respectively. Inclusion of diffuse basis
functions has been checked and shown unnecessary.
One important advantage coming from the localized nature of
the model is the linear scaling of the computational effort with the number of
atoms in the cluster. This has allowed us to perform full structural
relaxations of clusters with as many as 58 ions at a reasonable computational 
cost. Moreover, for each cluster size, a large number of isomers (between 10 and
15) has been investigated. The generation of the initial cluster geometries was
accomplished by using a pair potential, as explained in previous publications.
\cite{Agu99,Agu00b} The optimization of the geometries has been performed by 
using a downhill simplex algorithm.\cite{Pre91}

\section{Results and Discussion}

\subsection{Lowest Energy Structures of [(NaCl)$_m$(Na)$_2$]$^{2+}$ Cluster Ions}

In Fig.1 we present the optimized aiPI structures of the ground
state (GS) and lowest lying isomers or [(NaCl)$_m$(Na)$_2$]$^{2+}$ ($m$=8--28)
cluster ions. Below each
isomer we show the energy difference (in eV) with respect to the ground state.
The GS structures for $m$=6 and 7 are not shown in the figure because they were
calculated just to show the special stability of the $m$=8 size (see next
section). Nevertheless, they can be obtained by simply removing one and two
NaCl molecules from the GS structure for $m$=8. All the low-lying isomers of 
$m$=8 are based on the 3$\times$2$\times$2 structure of the (NaCl)$_6$ neutral 
cluster (where the notation indicates the number of ions along each of the three
perpendicular cartesian axes), and just differ in the way the six extra ions
are added to it. The most favorable location for these extra ions is along
two opposite edges of the neutral, so that the two triatomic NaClNa$^+$ units
minimice their mutual repulsion without distorting too much the structure of
the neutral cluster. Zhang and Cook have visualized this structure as the
combination of two 3$\times$3$\times$1 planar sheets.
\cite{Zha00} Given the bending of these sheets observed in the {\em ab initio}
calculations, we prefer to use the notation 3$\times$2$\times$2+3+3 for
this cluster, and will do it for the similar structures along this paper. 
Irrespective of the notation used, however, we must point out that our 
calculations essentially agree with their suggestion. The GS structures of
$m$=9 and 10 are much more distorted and difficult to visualize, but simply
result from the addition of one and two NaCl molecules, respectively, to the GS
isomer of $m$=8. For $m$=11, it is possible again to construct a quite compact
GS isomer by forming a 3$\times$3$\times$2+3+3 structure, again in good 
agreement with the suggestions of Zhang and Cook.\cite{Zha00} This time the two 
added triatomics are on the same face of the neutral structure,
due to the specific disposition of the ionic charges, and thus the screening
of the excess charge is not as complete as for the $m$=8 case. For $m$=12 a 
specially compact structure of a different kind appears. The GS for this size
can be obtained from that of the singly charged (NaCl)$_{13}$Na$^+$ cluster ion
by removing the inner six-coordinated chloride anion. This structure had been
suggested by the pair potential model calculations of Martin\cite{Mar82} and
by the experiments of Li and Whetten\cite{Li92} and Zhang and Cooks.
\cite{Zha00} These last authors use the term ``defect structure'' to refer to
this kind of structure, and in this case we will use the same notation.

These two kinds of structures seem to have a very high stability in the
whole size range considered in this study. For example,
a$\times$b$\times$c+3+3 fragments are observed for $m$=8,11,14,20 and 26. Defect
structures are adopted as GS structures for $m$=12 and 21. Note that for this 
last size the anion vacancy is not located in the center of the cluster but on 
an edge position. Anions are more stable the larger their coordination number
(the opposite holds for cations)\cite{Agu97a,Agu97b}, so the removal of an anion
with six coordination will be in general not favored energetically. $m$=12 is an
exceptional case in the sense that a highly compact and symmetrical structure
can be obtained by removing the central anion from a 3$\times$3$\times$3
singly charged cluster ion. A symmetrical structure tends to be favored by the
Madelung energy component, which is the most important contribution to binding
in ionic systems, and this compensates for the loss of the most stable anion
in the cluster. The same will not be true for most of the other values of $m$
where a defect structure can be formed. Another specially compact cluster
that could fit into the defect structure category is $m$=24, which can be 
obtained from the 4$\times$4$\times$3+3 structure of (NaCl)$_{25}$Na$^+$ by 
removing a corner anion. The only compact cluster that does not fit into any of 
these two categories is $m$=17, that can be viewed as the combination of two 
singly charged blocks, namely 3$\times$3$\times$3 and 3$\times$3$\times$1. 
Although this last structure coincides also with that advanced by Zhang and
Cooks,\cite{Zha00} a detailed comparison with their suggestions
shows that the agreement is not completely good for other sizes. For example, 
the GS structure of $m$=20 is predicted to result from the merging of two
3$\times$3$\times$3 blocks. We obtain indeed this structure as a low lying
isomer (see Fig. 1), so that those authors were not too far from the real
answer. Similarly, the GS structure for $m$=26 was predicted to be a combination
of 5$\times$3$\times$3 and 3$\times$3$\times$1 blocks instead of the structure
shown in Fig. 1.

The GS structures for the rest of the sizes are mainly obtained by adding or
removing NaCl molecules from the compact clusters of one of the two families
mentioned in the last paragraph. One exception could be $m$=23, which is formed
by adding a bent NaClNa$^+$ triatomic unit to a compact 5$\times$3$\times$3
structure.

Comparing to the results of our previous papers on neutral and singly charged
alkali halide clusters,\cite{Agu97a,Agu97b,Agu98} we appreciate that the
structures found in those cases can serve as ``seeds'' for the generation of
those of the doubly charged clusters. Specifically, the magic number
structures of the neutrals (AX)$_n$ ($n$=6, 9, 12, 15, etc) serve to generate
specially stable [(NaCl)$_m$(Na)$_2$]$^{2+}$ cluster ions with $m$=8, 11, 14,
17, 20, etc, by edge attaching of NaClNa$^+$ triatomic units. Specially compact 
doubly charged isomers can also be obtained by removing a chloride anion 
from one of the singly charged (AX)$_n$A$^+$ cluster ions, being this the case 
for $m$=12 and 21, or by adding a triatomic to the singly charged clusters, for
example $m$=17.

\subsection{Relative stabilities and connection to experimental mass spectra}

In the experimental mass spectra,\cite{Zha00} the populations observed for some 
cluster sizes are enhanced over those of the neighboring sizes. These ``magic 
numbers'' are a consequence of the evaporation/fragmentation
events that occur in the cluster beam, mostly after ionization.\cite{Ens83}
A magic cluster of size $m$ has a stability that is large compared to 
that of the neighboring sizes ($m$-1) and ($m$+1). Thus, on the average, 
clusters of size $m$ undergo a smaller number of evaporation/fragmentation 
events, and this leads to the maxima in the mass spectra. A most
convenient quantity to compare with experiment is the second energy difference
\begin{equation}
\Delta_2(m) = [E(m+1) + E(m-1)] - 2E(m),
\end{equation}
where E($m$) is the total energy of the [(NaCl)$_m$(Na)$_2$]$^{2+}$ cluster ion.
A positive value of $\Delta_2(m)$ indicates that the $m$-stability is larger
than the average of the ($m$+1)- and ($m$-1)-stabilities.

We show in figure 2 our results concerning the stabilities of the doubly
charged cluster ions. The magic numbers can be divided into two subsets:
sizes $m$=8,11,14,17,20 and 26 show large maxima in the $\Delta_2(m)$ curve;
sizes $m$=9,12,21 and 24 show smaller but positive values of $\Delta_2(m)$. All
the enhanced stabilities found in the experiments in this size range, namely
$m$=11,12,17,20,21 and 26,\cite{Zha00} are reproduced by our calculations. 
Sizes $m$=8 and 9 are too small to be observed in the experiments
by Zhang and Cooks, who found a critical size for the stability of the
doubly charged cluster ions of $m$=11. After the stability island found at sizes
$m$=11--12, no doubly charged cluster ion was observed in the experiments
\cite{Zha00} until reaching a value of $m$=17, so that the magic number $m$=14
is not observed either. Finally, although $m$=24 is not considered a magic
number in their paper\cite{Zha00} due to some scatter in the experimental data,
it is concluded that it might exhibit some enhanced stability. Thus, the
agreement with experiment can be considered excellent. It is a very interesting
question that deserves further investigation, however, why the metastable
potential energy minima of cluster ions in the size range $m$=13--16 can not be
populated in the experiments. Li and Whetten\cite{Li92} produced the doubly
charged cluster series by soft anion photoejection from the singly charged
(AX)$_n$A$^+$ series, and found that the stability island observed for 
$m$=11--12 is a consequence of the high efficiency of that process for the 
parent cluster with $n$=13. We note that halogen photoejection from the GS 
structure of (AX)$_{14}$A$^+$ found in previous publications\cite{Agu98} would 
lead directly to the GS isomer of [(NaCl)$_{13}$(Na)$_2$]$^{2+}$ shown in Fig. 
1, but in this case the process is not so efficient as for $n$=13. One could 
speculate that the one-coordinated cation left in the structure is very prone to
dissociate even for very modest excess vibrational energies. On the other hand, 
photoejection of a halide anion from the GS structure of (AX)$_{15}$A$^+$,
which is also based on attaching ions to a 3$\times$3$\times$3 compact cube,
would not directly populate the GS structure of [(NaCl)$_{14}$(Na)$_2$]$^{2+}$,
which is an elongated structure. Nevertheless, calculations on the evaporation
kinetics processes would be needed in order to draw definite conclusions.

Now we try a rationalization of the enhanced stabilities in terms of structural
properties. Hopefully, this will allow the GS structures of clusters larger
than those explicitely included here to be predicted with some confidence.
A general feature of [(NaCl)$_m$(Na)$_2$]$^{2+}$ cluster ions in the size
range considered in this paper is that a$\times$b$\times$c+3+3 fragments are 
specially stable compared to other isomers whenever they can be formed. The
apparent reason is that those structures tend to minimice the repulsion between
the two excess positive charges while not distorting too much the compact
a$\times$b$\times$c structures of the neutrals, which are energetically favored
by purely Madelung energy considerations.\cite{Agu97a,Agu97b} As the prefered
place to attach the NaClNa$^+$ triatomic units is along edges of the neutral
structures, for larger sizes (where none of the three edges will contain just 
three ions) one can advance a corresponding relevance of a$\times$b$\times$c+5+5
structures. In all cases, at least one of the three edges of the neutral
structures has to contain an even number of ions in order to preserve charge
neutrality and expose a
convenient binding site for the tri- or penta-atomic chains. To these structural
families we have to add the defect structures obtained by removing a halide
anion from the a$\times$b$\times$c compact structures that occur for the singly 
charged (AX)$_n$A$^+$ cluster ions when all three edges contain an odd number
of ions. In Table I we show all the relevant fragments of those kinds.
Each series (except the defect one) has a typical periodicity that
could in principle be reflected in different portions of the mass spectra, given
the high stability of these fragments. We can see that the great majority of
the magic numbers observed in the experiments by Zhang and Cooks\cite{Zha00}
can be explained in terms of the structures shown in the table. Thus, 
$m$=12,21,(24),30,36,52 and 61 are ascribed to defect structures based on
3$\times$3$\times$3, 5$\times$3$\times$3, (4$\times$4$\times$3+3), 
7$\times$3$\times$3, 5$\times$5$\times$3, 7$\times$5$\times$3 and 
5$\times$5$\times$5 parent singly charged structures 
($m$=52 was observed to show some enhanced stability,
although less than those of the others).\cite{Zha00} $m$=8,11 and 14 are
p$\times$2$\times$3+3+3 structures with p=2,3 and 4, respectively. $m$=20,26
and 32 are p$\times$4$\times$3+3+3 with p=3,4 and 5. Finally, m=34,44,54 and 64
are p$\times$4$\times$5+5+5 structures with p=3,4,5 and 6. Note that the values
of $p$ are as close as possible to the lengths of the other edges, as expected. 
With the only exception of $m$=17, all the experimental magic numbers fit into
any of these categories, which we consider evidence enough for the 
correcteness of the structures proposed. We note that the stability of the
defect-like structures is reduced with respect to that shown by the other magic
numbers (see Fig. 2). Note also that the inclusion of the
merged block structures of Zhang and Cooks\cite{Zha00} is not needed to
explain the enhanced stabilities, even though they will surely be low energy
isomers for those sizes where they can be formed.

\section{Conclusions}

The structures and stabilities of doubly charged [(NaCl)$_m$(Na)$_2$]$^{2+}$
cluster ions have been studied in the size range $m$=6--28 by means of {\em ab 
initio} Perturbed Ion calculations. For this size range, we have found two
main groups of specially compact structures: (a) those obtained by adding
triatomic chains to the edges of a$\times$b$\times$3 perfect neutral cuboids
(NaCl)$_n$, with $n$=$m$-2,
and (b) those obtained by removing one chloride anion from the perfect
a$\times$b$\times$c cuboids of the singly charged cluster ions
(NaCl)$_n$Na$^+$, with $n$=$m$+1. The way in which these structures are 
constructed indicates that there is a correlation with the structures found 
previously for neutral and singly charged alkali halide clusters.
\cite{Agu97a,Agu97b,Agu98} A comparison with the structural assignment
suggested by Zhang and Cooks\cite{Zha00} after an interpretation of their
collision induced fragmentation spectra shows a good level of agreement. 
Nevertheless, there are some minor discrepancies. For example, the merged
block structures suggested in their work are not found to be the GS isomers
for any size, even though they are low lying structural isomers.

The calculated enhanced stabilities show an excellent agreement with the
experimental results, being this an ideal check for the correctness of our
theoretical calculations. The only calculated magic numbers that are not
present in the experimental mass spectra are those of $m$=8, 9 and 14. The
experiments are not able to populate the metastable potential energy minima
of these structures, however, so that no comparison is possible in these cases.
With the only exception of $m$=17, we find that all the enhanced cluster
stabilities are a consequence of the highly compact structures that can be 
built for certain values of $m$ and mentioned in the previous paragraph.
Given the high stability of the structures obtained by adding triatomic chains
to compact neutral structures, we have proposed that the structures resulting
from adding pentaatomic chains to the edges of larger neutral clusters should
also be specially stable. Taking all these structural families altogether, we
have shown that we can reproduce all the magic numbers observed in the 
experimental mass spectra.

$\;$

$\;$

{\bf ACKNOWLEDGEMENTS:} 
The author is grateful to the Ministerio de Educaci\'on y Ciencia of Spain for
the concesion of a postdoctoral grant.

\newpage

{\bf Captions of Figures and Tables.}

{\bf Figure 1}. Lowest-energy structure and low-lying isomers of
[(NaCl)$_m$(Na)$_2$]$^{2+}$ cluster ions. Dark balls are Na$^+$ cations and 
light balls are Cl$^-$ anions. The energy
difference (in eV) with respect to the most stable structure is given below
the corresponding isomers.

{\bf Figure 2}.
Size evolution of $\Delta_2$(m) (eq. 1).
The local maxima in the second energy difference curve are shown explicitely.

{\bf Table I}
Structural series, together with their inherent periodicities, used to explain
the experimentally observed magic numbers. Those cluster sizes $m$ 
that are actually observed to show an enhanced stability in the mass spectra of 
[(NaCl)$_m$(Na)$_2$]$^{2+}$ clusters are written in boldface. $m$=17 is the
only exception (see text).

\begin {table}
\begin {center}

\begin {tabular} {|c|c|c|} \hline
 Structure & Periodicity & Cluster size n \\
\hline
p$\times$2$\times$3+3+3 & 3 & {\bf 8,11,14},17,... \\
p$\times$4$\times$3+3+3 & 6 & {\bf 14,20,26,32},38... \\
p$\times$4$\times$5+5+5 & 10 & 24,{\bf 34,44,54},64... \\
a$\times$b$\times$c-1 & -- & {\bf 12,21,(24),30,36,52,61},... \\
\end {tabular}
\end {center}
\end {table}


%
%
%
%
%
%
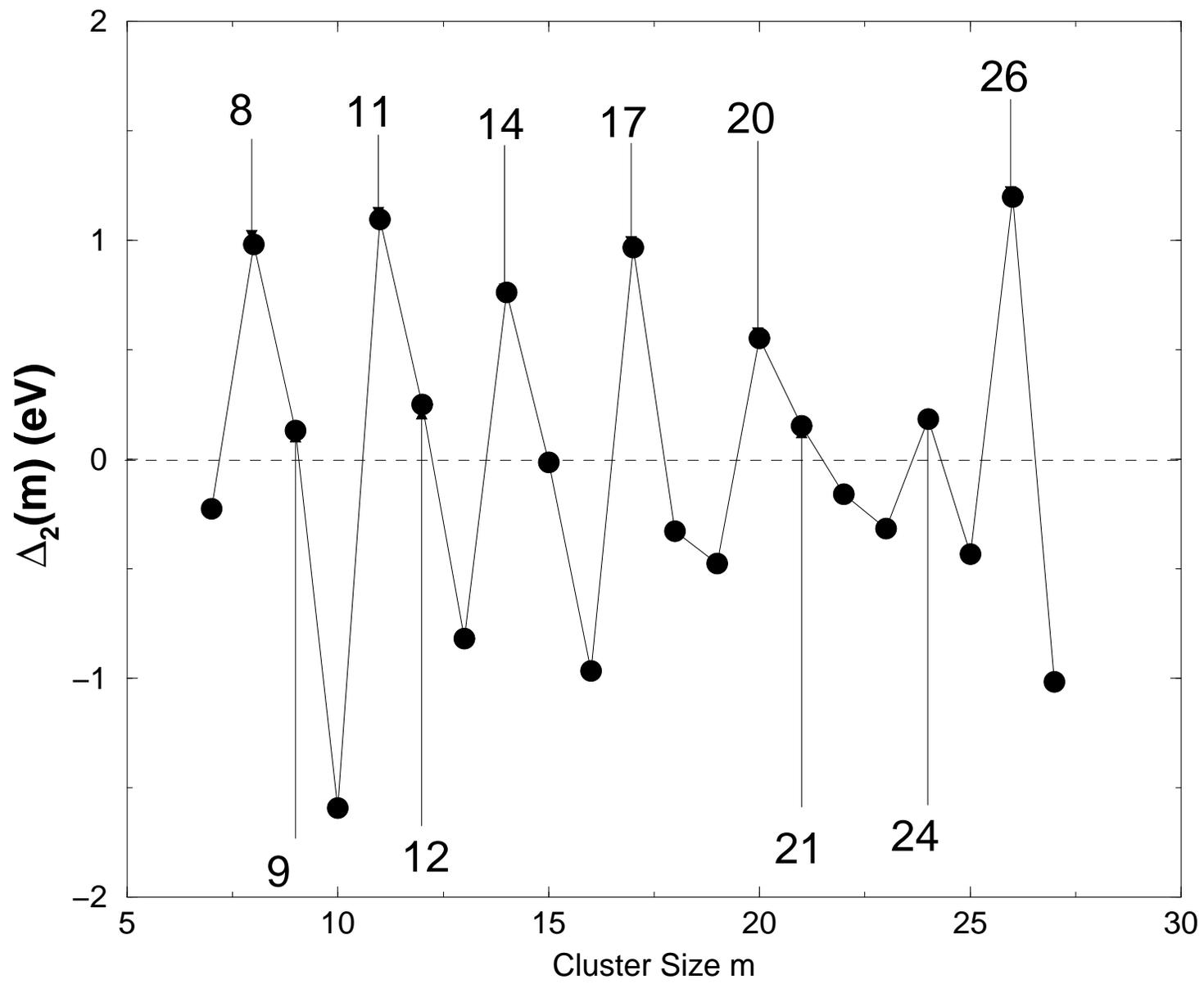
\end{document}